\documentclass{aa}
\usepackage{graphicx}
\usepackage{lscape}
\usepackage{subfigure}

\usepackage{txfonts}
\begin{document}
\title{Rotation-disk connection for very low mass and substellar objects in the Orion Nebula Cluster\thanks{Full Table\,2 is only available in electronic form at the CDS via anonymous ftp to cdsarc.u-strasbg.fr (130.79.128.5) or via http://cdsweb.u-strasbg.fr/cgi-bin/qcat?J/A+A/}}

 	\author{Mar\'{i}a V. Rodr\'{i}guez-Ledesma
          \inst{1,}\inst{2,}\inst{3},         
          Reinhard Mundt\inst{1},
\and
          Jochen Eisl\"offel\inst{3}
	 }
  \institute{Max-Planck-Institut f\"ur Astronomie (MPIA), K\"onigstuhl 17, D-69117 Heidelberg, Germany\\
              \email{vicrodriguez@mpia.de}
	     \and
International Max Planck Research School for Astronomy \& Cosmic Physics at the University of Heidelberg
\and
Institut f\"ur Astrophysik, Georg-August-Universit\"at, Friedrich-Hund-Platz 1, D-37077 G\"ottingen, Germany
\and
             Th\"uringer Landessternwarte Tautenburg, Sternwarte 5, D-07778 Tautenburg, Germany
             }


 
  \abstract
   {}
{Angular momentum (J) loss requires magnetic interaction between the forming star and both the circumstellar disk and the magnetically driven outflows. In order to test these predictions many authors have investigated a rotation-disk connection in pre-main sequence objects with masses larger than about 0.4\,M$_{\odot}$. For brown dwarfs (BDs) this connection was not investigated as yet because there are very few samples available. We aim to extend this investigation well down into the substellar regime for our large sample of $\approx$\,80 BDs in the Orion Nebula Cluster, for which we have recently measured rotational periods.}
{In order to investigate a rotation-disk correlation, we derived near-infrared (NIR) excesses for a sample of 732 periodic variables in the Orion Nebula Cluster with masses ranging between $\approx$\,1.5-0.02\,$M_\odot$ and whose IJHK colors are available. Circumstellar NIR excesses were derived from the $\Delta$[I-K] index. We performed our analysis in three mass bins ($>$\,0.4\,$M_{\odot}$, 0.4-0.075\,$M_{\odot}$, and $<$\,0.075\,$M_{\odot}$).}
{We found a rotation-disk correlation in the high and intermediate mass regime, in which objects with NIR excess tend to rotate slower than objects without NIR excess. Interestingly, we found no correlation in the substellar regime. A tight correlation between the peak-to-peak (ptp) amplitude of the rotational modulation and the NIR excess was found however for all objects with available ptp values ($<$\,0.4\,$M_{\odot}$). We discuss possible scenarios which may explain the lack of rotation-disk connection in the substellar mass regime. One possible reason could be the strong dependence of the mass accretion rate $\dot{M}$ on stellar mass in the investigated mass range ($\dot{M}$\,$\varpropto$\,$M^{2\,-\,2.8}_{star}$), which is expected to result in a corresponding mass dependence of $\dot{J}/J$.}
   {}
\titlerunning{Rotation-disk connection in the ONC}
\authorrunning{Rodr\'iguez-Ledesma et al.} 
 \maketitle
%
\section{Introduction}

At present observations and theory together give a comprehensive picture of star formation. Nevertheless, how young pre-main sequence (PMS) objects effectively lose angular momentum is still not fully understood. PMS stars increase their mass by accreting material from the circumstellar disk. At this stage, accretion and ongoing contraction would contribute to a net spinning up of the PMS star, which is expecting to rotate at nearly breakup velocity (e.g. Vogel \& Kuhi, \cite{Vogel1981}). Yet, numerous observations reveal that these objects rotate much slower, with typical rotational velocities of $\approx$10\% the breakup velocity (e.g. Herbst et al. \cite{H2007} and references therein). It is clear that considerable amounts of angular momentum are somehow removed from the star/disk system during the accretion phase. \\
Various theories, all of them requiring ongoing mass accretion from the disk to the star in combination with magnetically driven outflows, are usually proposed to explain this enormous removal of angular momentum and the large spread of rotational velocities observed (e.g. Hillenbrand \cite{Hill97}; Herbst et al. \cite{H2002} hereafter H2002; Lamm et al. \cite{Lamm04}; \cite{Lamm05}; Scholz \& Eisl\"offel \cite{Scholz04a}; \cite{Scholz05}; Rodriguez-Ledesma et al. \cite{Rodriguez09a} hereafter Paper\,1). The proposed theories rely on a disk-locking mechanism (e.g. Camenzind \cite{Camenzind1990}; K\"onigl \cite{Koenigl1991}), X-winds (e.g. Shu et al. \cite{Shu1994}; Ferreira et al. \cite{Ferreira06}), disk winds or massive stellar winds working in conjunction with magnetospheric accretion (e.g. Matt \& Pudritz, \cite{Matt2005b}) to account for the angular momentum loss. Irrespective of the specific model, there is a general consensus that circumstellar accretion disks are somehow responsible for the removal of angular momentum, and therefore accreting stars should on average rotate slower than non-accreting ones.\\
There is wide observational evidence that most low mass pre-main sequence stars and even substellar objects have circumstellar disks (e.g. Hillenbrand et al. \cite{Hill98}; Muench et al. \cite{Muench2001}; Mohanty et al. \cite{Mohanty}; Rebull et al. \cite{Rebull2006}; etc).  Accretion features have also been observed in substellar objects (e.g. Mohanty et al. \cite{Mohanty}). \\
For more than 15 years, several authors investigated possible correlations between rotation rates and the presence of circumstellar accretion disks (e.g. Edwards et al. \cite{Edwards}; Rebull, \cite{Rebull2001}; H2002; Lamm et al. \cite{Lamm05}; Rebull et al. \cite{Rebull2006}; Cieza and Baliber, \cite{Cieza2007}). While many of them found a rotation-disk correlation for solar and low mass stars (e.g. H2002; Lamm et al. \cite{Lamm05}; Rebull et al. \cite{Rebull2006}; Cieza and Baliber, \cite{Cieza2007}; Biazzo, \cite{Biazzo2009}), others did not (e.g. Rebull, \cite{Rebull2001}; Makidon, \cite{Makidon2004}; Nguyen et al. \cite{Nguyen2009}). As stated by Cieza \& Baliber (\cite{Cieza2007}), intrinsic dependence of rotation on mass and age, which were not taken into account in many of the previous works, were probably one of the reasons for the discrepant results. An additional reason is probably the use of different and sometimes ambiguous accretion indicators. Most of the above mentioned studies are using near-infrared (NIR) excess as a circumstellar disk indicator. While NIR excess probes the presence of circumstellar dust, it does not probe ongoing accretion. However, a remarkable correlation between mass accretion indicators such as H$\alpha$ emission line widths and NIR excess has been reported in the literature (e.g. Sicilia-Aguilar et al. \cite{Sicilia2005}; \cite{Sicilia2006}) and therefore using NIR excess as a mass accretion indicator is a reasonable approach.\\ 
Muench et al. (\cite{Muench2001}) found that about 50\% of their brown dwarf candidates in the Trapezium region show near-infrared (NIR) excesses indicative of a circumstellar disk. Lada (\cite{Lada2000}) found that $\approx$50\% of objects in the Trapezium cluster have a JHK excess, while their observations at $\approx$3$\mu$m revealed a much higher disk fraction of about 85\%. Since longer wavelengths provide a better disk indication than observations at NIR wavelengths, the use of Spitzer data (IRAC bands: 3.6\,$\mu$m, 4.5\,$\mu$m, 5.8\,$\mu$m, and 8\,$\mu$m) seems to be the solution towards an unambiguous disk indicator. However, the evaluation of the IRAC data of the ONC encounters two serious problems, namely a very strong nebular background and high stellar densities. These two problems obviously get stronger with decreasing distance to the cluster center. As stated by Rebull et al. (\cite{Rebull2006}), due to the strong nebular background and source confussion, even relatively bright periodic variables close to the Trapezium are missing IRAC detections. Megeath et al. (\cite{Megeath2005}) estimated that about 700 stars may be missing IRAC detection in the inner 5\,x\,5\,arcmin$^2$ center of the ONC. Only objects in the ONC with spectral types earlier than M2 (i.e. M\,$\gtrapprox$\,0.4\,$M_\odot$ if Baraffe et al. (\cite{Baraffe98}) models are used) were analyzed and searched for the presence of a disk and a rotational-disk connection by means of IRAC data (Rebull et al. \cite{Rebull2006} and Cieza \& Baliber \cite{Cieza2007}). Since our sample of periodic variables is three magnitudes deeper than previous studies and contains many objects in the inner most part of the ONC, we have to rely on NIR data to identify the circumstellar disks in our sample. Although NIR observations are less efficient in detecting disks than observations at longer wavelengths, the relatively high disk fraction of $\approx$50\% found by Lada (\cite{Lada2000}) and Muench et al. (\cite{Muench2001}) among large samples of low mass stars and brown dwarfs by means of JHK excesses provides valid support that NIR excess is a suitable disk indicator for low mass and substellar objects.\\
A considerable number of studies, which have explored the connection between the presence of disks and rotation in low and very low mass stars, are summarized in Herbst et al. \cite{H2007} and Bouvier, \cite{Bouvier07}. The major aim of this study is to constrain for the first time a possible rotation-disk connection in the substellar mass regime, and compare these results with higher mass objects. This study would not have been possible without the rotational period measurements of 124 BD candidates in Paper 1.\\
In Paper\,1 we confirmed the strong period dependence on stellar mass found for higher mass stars by H2002 and extended it into the brown dwarf regime down to $\approx$\,20\,M$_{jup}$. We found that substellar objects tend to rotate faster than their stellar counterparts. In order to constrain a possible rotation-disk connection, in particular for the so far very poorly investigated very low mass and substellar regime, we analyze here a sample of 732 periodic variables from Paper\,1 and Herbst et al. \cite{H2002} which have available JHK colors to be used as disk indicators. In Sect. 2 we give details of the used NIR data sets. Section\,3 describes how we estimated the masses in our sample. Section\,4 defines the procedure to account for objects with and without NIR excess. We present our results on the correlations between rotational periods and peak-to-peak amplitudes with NIR excess in Sect. 5 and 6 and briefly address the disk fraction found in Sect.7. In Sect. 8, we discuss in detail possible scenarios which can result in a lack of correlation between NIR and rotation and summarize our results in Sect. 9. 
\begin{figure}[t!]
\centering
\rotatebox{270}{\includegraphics[width=8.45cm, height=9cm]{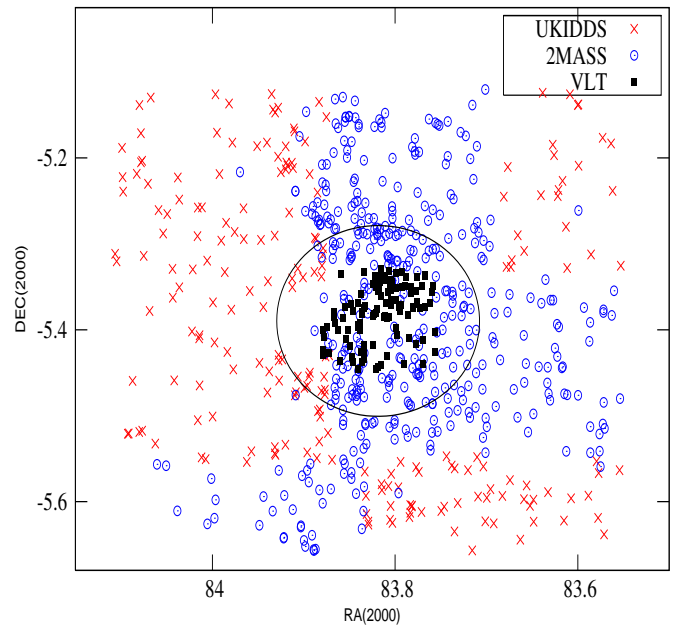}}\\
  \caption{Spatial distribution of all 732 periodic variables with near-infrared counterpart. Black squares are 113 objects with VLT JHK measurements, red crosses are the 210 periodic variables with UKIDSS JHK measurements and blue open circles are the 409 periodic variables with 2MASS measurements.  The circle centered on $\Theta^1$ Ori with a radius of $\sim$ 6\farcm7 represents the cluster radius (see Paper\,1 for details).}
         \label{Spatialdistri}
   \end{figure}

\section{Optical and near-infrared data}

The I band data were obtained with the wide field imager (WFI) at the MPG/ESO 2.2-m telescope on La Silla, Chile (see Paper\,1 for details). We used the combined (i.e. Paper\,1 and H2002) I band data, which include 746 periodic variables with masses ranging from about 1.5\,$M_\odot$ (H2002) down into the BD regime. We combined these optical data with JHK data from three different sources: UKIDSS\footnote{The UKIDSS project as defined in Lawrence et al \cite{Lawrence07} uses the UKIRT Wide Field Camera (WFCAM; Casali et al, \cite{Casali07}).}, VLT (MacCaughrean et al. in prep.), and 2MASS. Ideally we would have used only UKIDSS since it provides much deeper and much more accurate JHK data than 2MASS, but unfortunately only less than 50\% of our ONC WFI field (33 x 34 arcmin$^2$, center approximately at the position of $\Theta^{1}$ Ori) is covered in all three bands by the UKIDSS survey. In particular the innermost part of the ONC is not covered (for details see Fig.\,1 ). We therefore used for a fraction of the inner ONC region deep VLT-ISAAC data (MacCaughrean et al. in prep.) covering approximately the central 7\,x\,7\,arcmin$^2$. For those regions in our WFI field which were neither covered by UKIDSS nor by the VLT data we used 2MASS data. The cross-identification between the periodic variables and these near-infrared catalogues was done by looking for positional matches within a 2\arcsec search radius. Finally 732 out of the 746 known periodic variables from Paper\,1 have an infrared counterpart for which data in J, H, and K are available.\\
Many objects have JHK data in more than one of the used catalogues. Due to their higher accuracy we gave higher priority to the UKIDSS and VLT data than to the 2MASS data (see Sect. 2.2 for details).\\
The basis of the VLT catalogue are VLT-ISAAC images. This catalogue is supplemented and calibrated with data  
from a sizeable number of other surveys which have been transformed to the 2MASS system resulting in a homogeneous catalogue. To allow a proper comparison, we transformed the UKIDSS magnitudes into the 2MASS system using the relations described in Hewett et al. (\cite{Hewett2006}). The magnitude distribution of the periodic variables in I and JHK are shown in Fig.\,2 with separate distributions for the three infrared data sets used in this paper. As summarized in Table\,1, we finally used 210, 113, and 409 JHK measurements from UKIDSS, VLT and 2MASS, respectively. Most objects in the innermost part of the ONC, for which 2MASS data had to be used, are bright objects from H2002, 95\% of which have J magnitudes brighter than 14\,mag.
\begin{figure*}
\centering
\includegraphics[width=16cm,height=13cm]{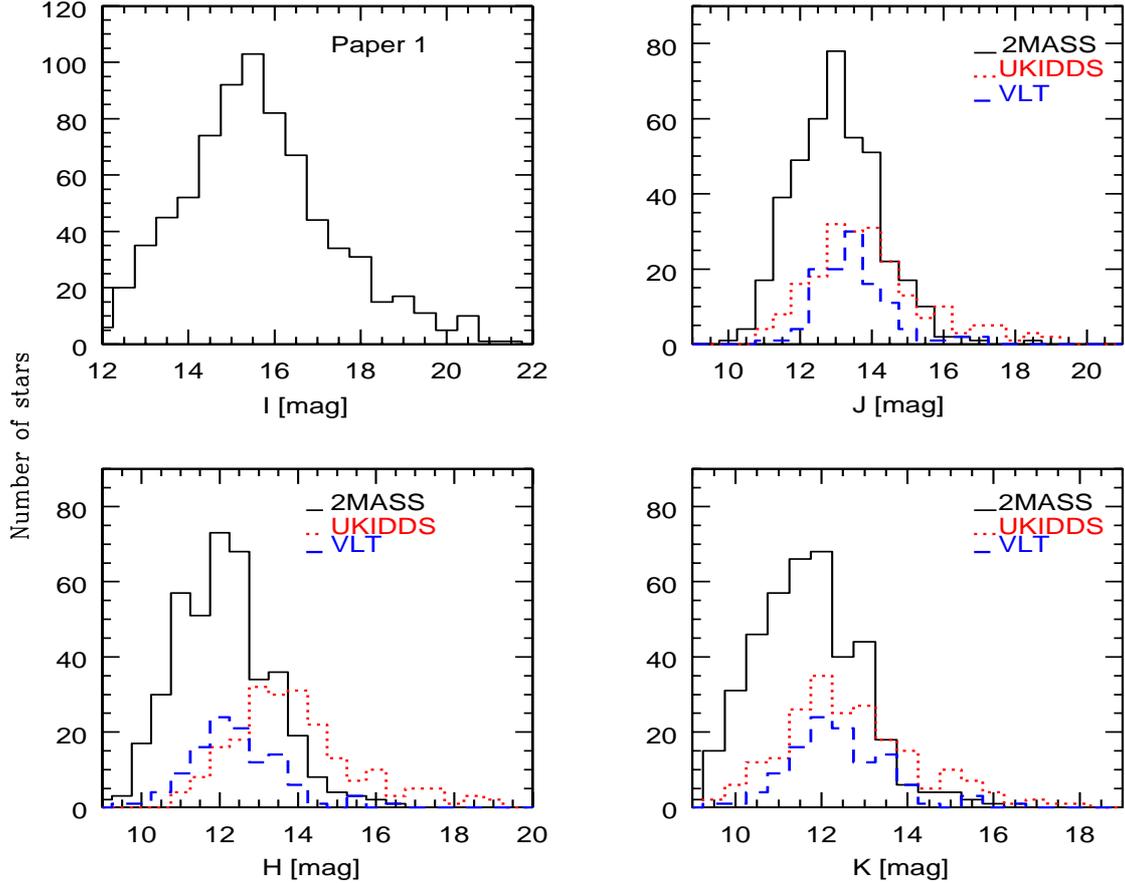}\\
\caption{Histograms showing the magnitude distribution of the periodic variables in I, J, H, and K. The three different NIR data sets finally used in this paper are shown separately. The I band data from Paper\,1 are shown for comparison purposes.}
         \label{IRbands}
   \end{figure*}

\subsection{Photometric errors of I-K, J-H and H-K colors}
As we will outline in Sect. 4, our IR excess determinations are based on I-K, J-H, and H-K colors. Therefore we have to briefly discuss the accuracy of these color determinations.\\
There are basically two sources of errors: photometric errors of the individual I, J, H, and K measurements and errors resulting from variability of the periodic variables in combination with the non-simultaneous data acquisition in the used bands.\\ Non-simultaneous data acquisition occurred for all I-K colors. As outlined in Paper\,1 (see there Fig.\,14), the peak-to-peak (ptp) amplitudes in the I band for periodic variables of 0.075-0.4\,$M_{\odot}$ have a median value of about 0.06\,mag, and for 85\% of them the ptp values were below 0.2\,mag. Since we will rarely observe the periodic variables in minimum in one band and in maximum in the other band, I-K colors are expected to have median errors due to variability of at most 0.06 mag. For the J-H and H-K colors, it is irrelevant in the 2MASS data, which are practically taken simultaneously. This means that only for the UKIDSS and VLT-ISAAC data one has to investigate the relevance of this effect.  Since the ptp amplitudes are expected to be significantly lower in the NIR than in the I band, the J-H and H-K colors calculated from the UKIDSS and VLT data are expected to have median errors due to variability of $\lesssim$\,0.05\,mag.\\
The expected mean photometric errors depend on the survey/catalogue used. I band photometry from Paper\,1 has typical photometric errors for a I=14, I=18 and I=20 of $\approx$\,0.001\,mag, $\approx$\,0.01\,mag, and $\approx$\,0.03\,mag, respectively. \\
The UKIDSS data are very deep and accurate with typical photometric errors for a J=14 , H=13.5 and/or K=13\,mag object of $\approx$\,0.002\,mag and typical errors of $\approx$\,0.01\,mag for 2\,mag fainter objects. As stated above, observations in the three bands are not taken simultaneously. Nevertheless, as already discussed, we do not expect variability to cause errors in extent of 0.05\,mag.\\
The 2MASS data have photometric errors which are on average one order of magnitude higher than those from UKIDSS for the same magnitude. As already stated above, objects for which 2MASS data had to be used are those located in regions of our WFI field that are not covered by neither UKIDSS nor VLT (see Fig.\,1).\\
The VLT catalogue reaches 5$\sigma$ point source limiting magnitudes at approximately 22, 21, and 20 for J,H, and K, respectively. This is about 2\,mag fainter than UKIDSS, and typical photometric errors are therefore expected to be $\approx$\,5 times smaller than those of UKIDSS for the same magnitude.  As already mentioned the VLT data were not taken simultaneously. However, each data point is the average of several observations separated over timescales of hours to years, and consequently ``variability errors'' will be smoothed out to some degree.\\
We expect uncertainties in the I-K color determination dominated by the non-simultaneous data acquisition of $\lesssim$\,10\%. From the typical photometric errors we expect uncertainties in the J-H and H-K colors calculation of about 1\% for the VLT data and for the 2MASS data between 3\% and 10\% for a K\,=\,12\,mag and K\,=\,14\,mag objects respectively. The colors derived from UKIDSS data are affected by both photometric and variability errors. The J-H and H-K color errors are expected to be clearly smaller than 5\% and 7\% for a K\,=\,13 and 15\,mag object, respectively. This means that in principle we are able to detect excesses from the intrinsic J-H and H-K colors which are larger than about 1-5\% for objects with masses of $\approx$\,0.1\,M$_{\odot}$ and 7-10\% for the less massive objects in our sample. A rough estimate of the amount of NIR excess under typical stellar\,+\,disk system parameters (e.g. Meyer et al. \cite{Meyer1997} and Hillenbrand et al. \cite{Hill98}) results in H and K band excesses which are expected to be of about 20\%. Although the amount of NIR excess is highly dependent on the different stellar\,+\,disk parameters, the estimated errors in the colors should not significantly affect our disk detectability. 
\begin{table}
\centering
\begin{minipage}[c]{\columnwidth}
\renewcommand{\arraystretch}{1.2}  
\caption{Number of periodic variables from Paper\,1 with a near-infrared counterpart}
\label{NIRcat}
\addtolength{\tabcolsep}{7pt}
\begin{tabular}{c|c|c} 
\hline\hline             
&\multicolumn{2}{c}{Periodic variables} \\
\hline
&At least one band & All JHK bands\\
\hline
UKIDSS & 419 & 210\\
VLT & 113 & 113\\
2MASS & 684 & 409\footnote{684 periodic variables in our sample have 2MASS JHK colors, but since UKIDSS and VLT data have higher priority than 2MASS, 2MASS colors have been used only for the 409 objects without UKIDSS and VLT data.}\\
\hline
\end{tabular}
\renewcommand{\footnoterule}{} 
\end{minipage}
\end{table}


\section{Mass estimates: Color-magnitude diagrams}

To estimate masses from evolutionary models we need to know individual extinction values, which requires the knowledge of the spectral type. Since the spectral types of most of our periodic variables are unknown, individual extinction values cannot be determined. For the sake of simplicity we assumed in Paper\,1 a uniform extinction value of A$_v$\,=\,1.4\, mag for all stars in the field, and we estimated masses based on the I band values. Although we still lack information on spectral types and therefore individual extinction values of our sample, we can make use of the near-infrared colors to perform additional/independent mass estimates, which are probably more accurate than the estimates based on the I band alone, due to the smaller extinction in the near-infrared bands and the smaller variability in this wavelength range. \\
We used the J vs I-J color-magnitude diagram for the mass estimates. In principle, H and K bands would be even less affected by extinction than the J band, but in these bands infrared excess due to stellar disks can pretend a much higher stellar flux than is actually present. We investigated the median extinction values for the inside and outside R$_\mathrm{cluster}$ in the literature (see Fig.\,\ref{Spatialdistri} and Paper\,1). First, we computed the median values of the extinction provided by Hillenbrand (\cite{Hill97}) in both spatial regions, which resulted in A$_v$=1.3\,mag and A$_v$=0.7\,mag for the inner and outer regions, respectively. Second, we checked extinction maps by Chaisson \& Dopita (\cite{Chaisson1977}). The latter reveals that there is no strong increase in the extinction towards the inner regions, with median extinction values of about 1 and 1.5 in the innermost parts of the ONC and below 1 in the outer regions. On the basis of these studies we assumed an average A$_v$ of 1.4\,mag for objects located inside R$_\mathrm{cluster}$ and 0.7 for objects outside R$_\mathrm{cluster}$. We used the assumed extinction values together with a 1\,Myr isochrone by Baraffe et al. (\cite{Baraffe98})\footnote{As in Paper\,1, we use models by Baraffe et al. (\cite{Baraffe98}) for our mass estimations. To allow a proper comparison with previous studies, each time a mass value is given here it is transfered into the mass scale of the mentioned model. Mass values estimated from other models are given explicitly only if needed.} to select objects with masses higher than 0.4\,M$_{\odot}$ (i.e. $\approx$\,M2 or earlier), between 0.075-0.4\,M$_{\odot}$ (i.e. $\approx$ M2 to M6), and $<$\,0.075\,M$_{\odot}$ (i.e. later than $\approx$\,M6). We further constrained our sample by removing \textit{a)} objects to the left (i.e. bluer) of the 5 Myr isochrone from Baraffe et al. (\cite{Baraffe98}), which are presumably background objects or in some cases young stars with strong scattered light emission from a circumstellar envelope or a nearly edge-on disk, and \textit{b)} those objects located in the top right-hand part of the J vs I-J color-magnitude diagram, which are presumably foreground objects with J and I-J colors that cannot be explained by any, even highly extincted, 1\,Myr old star at the distance of the ONC. Finally, our sample is reduced to 638 periodic variables (Table\,2).\\
For 81 periodic variables, 64 outside and 17 inside R$_\mathrm{cluster}$, we derived masses of $<$\,0.075\,M$_{\odot}$. This number is smaller than the 124 BD candidates found in Paper\,1 from I band data alone, resulting from the lower extinction in the near-infrared bands with the possible combination of higher variability in the I band. The determination of individual extinction values and spectral type will be necessary to unambiguously determine the masses of the sample.
\begin{table*}[t!]
\renewcommand{\arraystretch}{1.2}  
\begin{minipage}[]{2\columnwidth}
\caption{Sample of the NIR catalogue of periodic variables in Paper\,1. The complete catalogue is available electronically at the CDS.}
\label{PHOTtable}
\addtolength{\tabcolsep}{2pt}
\begin{tabular}{ccccccccccc} 
\hline\hline             
ID\footnote{Identification numbers from Rodriguez-Ledesma et al. (2009)} & I\footnote{I magnitudes from Rodriguez-Ledesma et al. (2009).}[mag] & I$_{err}$[mag] & J[mag] & J$_{err}$[mag]& H[mag] & H$_{err}$[mag]& K[mag] & K$_{err}$[mag] & $\Delta{(I-K)}$\footnote{ $\Delta{(I-K)}$\,$>$\,0.3 indicates the precense of a circumstellar disk.} & Flag\footnote{ This flag shows the origin of the NIR data. The possibilities are: 2MASS, UKIDSS, VLT. The VLT data set was not provided with individual photometric errors. A description of the estimated errors is available in Sect.\,2.}\\
\hline
1 & 15.535&  0.002&    13.540&   0.030&	12.820 &  0.040 &    12.510&   0.030&     -0.115&   2MASS\\
3 & 13.205&   0.001&    11.284&   0.001& 10.970 &  0.001 &   10.116 &  0.000&    0.109  &  UKIDSS\\
7 & 15.637&   0.002&    13.930&   0.030& 13.370 &  0.040 &   12.940 &  0.020 &    -0.013 &  2MASS\\
13 & 15.551 &  0.002&   14.002 &  0.003 & 13.374  & 0.003  &  13.136  & 0.003  &   -0.125  & UKIDSS\\
10384&	17.954&	0.054&	14.960&	-1.000&	14.160&	-1.000&	13.520&	-1.000&	0.114&	VLT\\
10387&	17.674&	0.011&	14.770&	0.020&	13.660&	0.030&	13.070&	0.020&	0.334&	2MASS\\
20323&	14.495&	0.001&	12.260&	0.020&	11.210&	0.030&	10.810&	0.020&	0.145&	2MASS\\
20325&	15.397&	0.005&	13.140&	0.040&	12.410&	0.040&	12.100&	0.040&	-0.313&	2MASS\\
20326&	18.853&	0.047&	14.680&	-1.000&	13.270&	-1.000&	12.370&	-1.000&	1.683&	VLT\\
20672&	19.654&	0.048&	17.276&	0.022&	15.957&	0.023&	15.638&	0.022&	0.516&	UKIDSS\\
\hline
\end{tabular}
\renewcommand{\footnoterule}{} 
\end{minipage}
\end{table*}
\section{NIR excess determination}
\subsection{NIR excess from (I-K) colors}

We used an index based on (I-K) colors in which I-band fluxes are thought to be dominated by photospheric emission, while K-band fluxes can have an additional strong component from the circumstellar disk (Hillenbrand et al. \cite{Hill98}). This approach has the advantage of a longer wavelength base, which should in principle result in more pronounced excess than other commonly used indices (e.g. $\Delta$(H-K)). Unfortunately, a NIR excess derived from I-K colors requires the knowledge of individual extinction values and spectral types, both unknown for most of our objects. As stated above, we assumed average A$_v$ values of 1.4\,mag and 0.7\,mag for objects inside and outside R$_\mathrm{cluster}$, following the methodology stated in the previous section. We are aware that near the ONC center many objects may be more highly extincted than the average, resulting in a sample of objects pretending to have disks, which is actually contaminated by highly extincted objects. On the other hand, in the outer regions a certain fraction of objects have lower extinction than the median value considered here and therefore we may underestimate the fraction of objects showing indications of a circumstellar disk in this region. Since we do not have information on individual spectral types, we need to make an assumption on the intrinsic photospheric color $(I-K)_0$. We used the observed I-J colors corrected by the assumed average A$_v$ values, and use the Baraffe et al. (\cite{Baraffe98}) color information for 1\,Myr old objects to derive individual intrinsic $(I-K)_0$ colors for our sample.\\
We then used the definition by Hillenbrand et al. (\cite{Hill98}) to account for the amount of near-infrared excess from the I-K color:
\begin{equation}
\Delta(I-K)=(I-K)_{measured}-0.5xA_v -(I-K)_{0},\\
\end{equation}
where $(I-K)_0$ are the intrinsic colors. \\
We considered $\Delta$(I-K)=0.3\,mag (Hillenbrand et al. \cite{Hill98}) as a lower limit in order to be regarded as a result from a circumstellar disk. We found that about 28\% of the whole sample show a NIR excess according to the $\Delta$(I-K) index. We denoted these objects as $\Delta$(I-K)+ in Table\,3.
Although several assumptions were made to calculate the $\Delta$(I-K) index, the values derived here will allow for the first time the comparison of the rotation-disk behavior between the higher mass objects and the very low mass and even substellar objects.

\subsection{Extinction-free indices}
For comparison purposes we also calculated extinction-free indices. The (unknown) extinction of the individual objects and the IR excess from a circumstellar disk cannot be easily disentangled in color-color diagrams. As a second disk-indicator, we use an ``extinction-free'' \footnote{To compute those indices, an extinction law has to be assumed and therefore they are not completely independent of extinction. In all cases we assumed the Rieke \& Lebofsky (\cite{Rieke1985}) extinction law with \textit{R=}\,3.1.} index Q as defined by Damiani et al. (\cite{Damiani2006}). In a J-H versus H-K diagram, there is a permitted strip of colors on which dwarfs should be located and which depends on the spectral type and the optical extinction. Since our sample includes objects in a broad mass range (i.e. from about 1.5\,M$_{\odot}$ to $\sim$\,20\,$M_{jup}$), we defined three limiting values of the index Q$_{JHHK}$ according to the three different mass regimes considered here. The extinction-free index provides a measurement of the deviation from these permitted strips. We then used the NIR index Q$_{JHHK}$ defined as follows
\begin{equation}
Q_{JHHK}=(J-H)-(H-K)(E(J-H)/E(H-K)),\\
\end{equation}
in which the ratio of the color-extinctions E(J-H)\,/\,E(H-K) is based on the extinction law of Rieke \& Lebofsky (\cite{Rieke1985}).\\
We note that the calculated Q$_{JHHK}$ values provide a conservative limit on the presence of an IR excess and therefore the presence of a disk. One should keep in mind that many objects in the permitted strip presumably show red J-H and/or H-K colors due to a combination of extinction \textit{and} near-infrared emission from a circumstellar disk (see Meyer et al. \cite{Meyer1997}). For many objects located at larger distances from the central Trapezium cluster, where the average extinction is expected to be relatively small, NIR colors redder than those of dwarf stars are more likely to be due to a circumstellar disk, and are missed by the Q$_{JHHK}$. This conservative disk indicator explains the $\sim$20\% disk fraction found, a value significantly smaller than the one derived from the $\Delta$(I-K) index. Although the Q$_{JHHK}$ index misses a large fraction of disks, we will mention the results obtained from it for comparison purposes.\\

\begin{figure}[t!]
\centering
\includegraphics[trim = 0mm 0mm 27mm 0mm,clip,width=8.5cm, height=6.73cm]{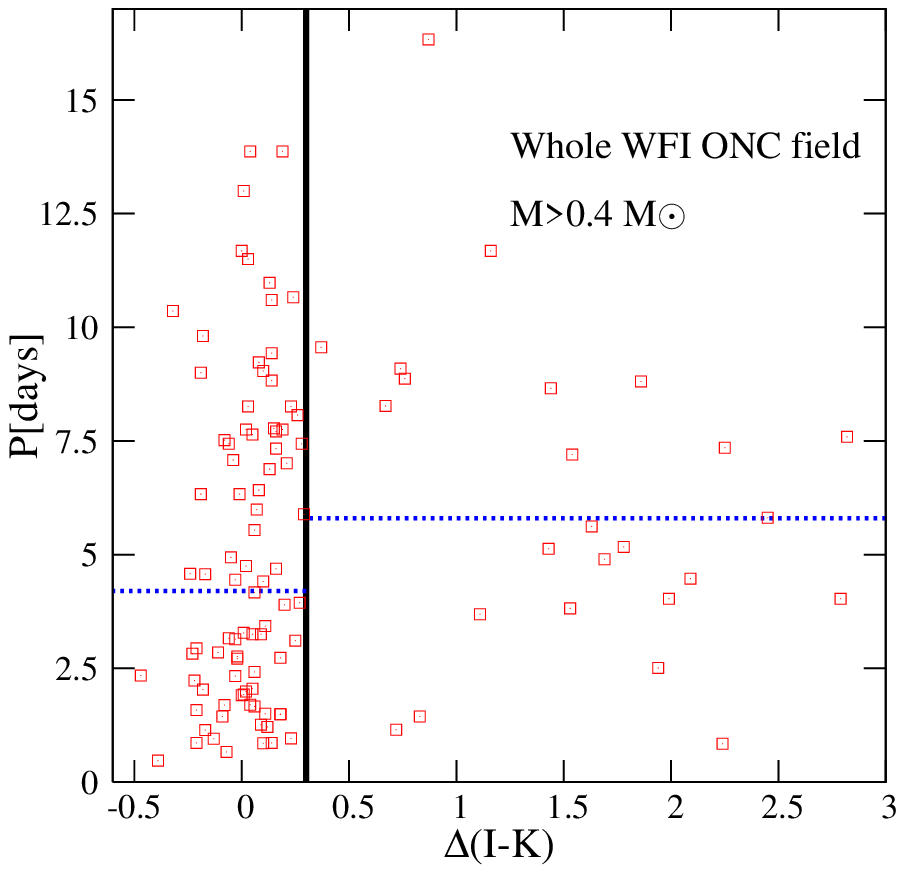}\\
\includegraphics[trim = 0mm 0mm 27mm 0mm,clip,width=8.5cm, height=6.73cm]{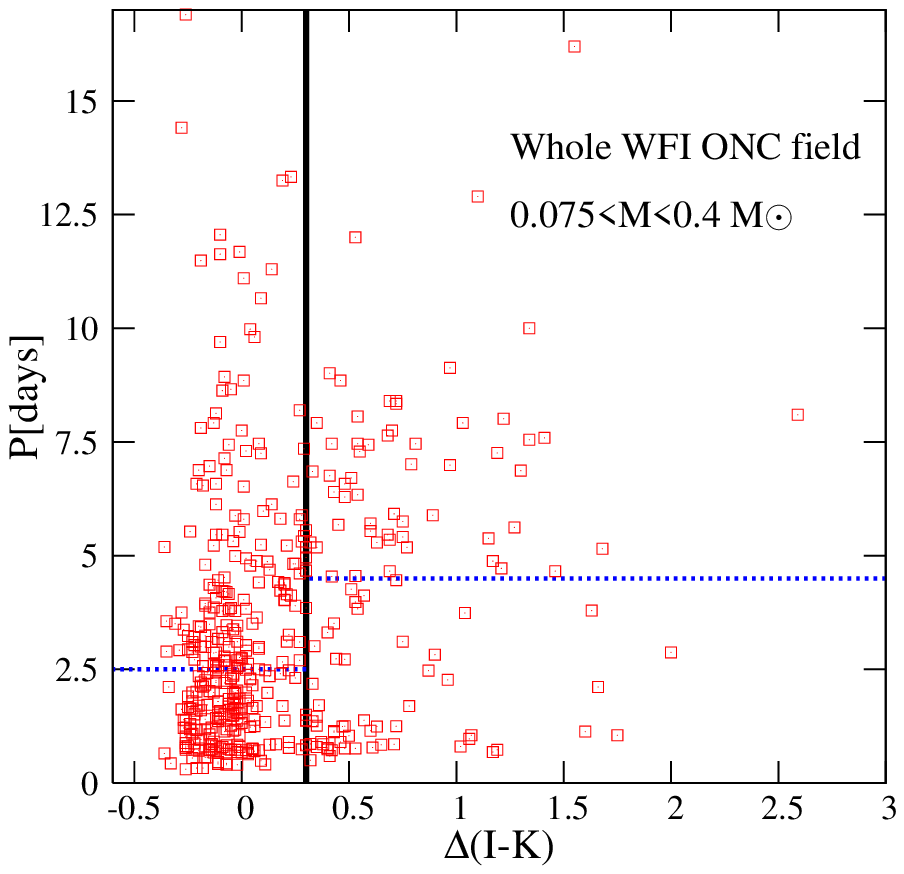}\\
\includegraphics[trim = 0mm 0mm 27mm 0mm,clip,width=8.5cm, height=6.73cm]{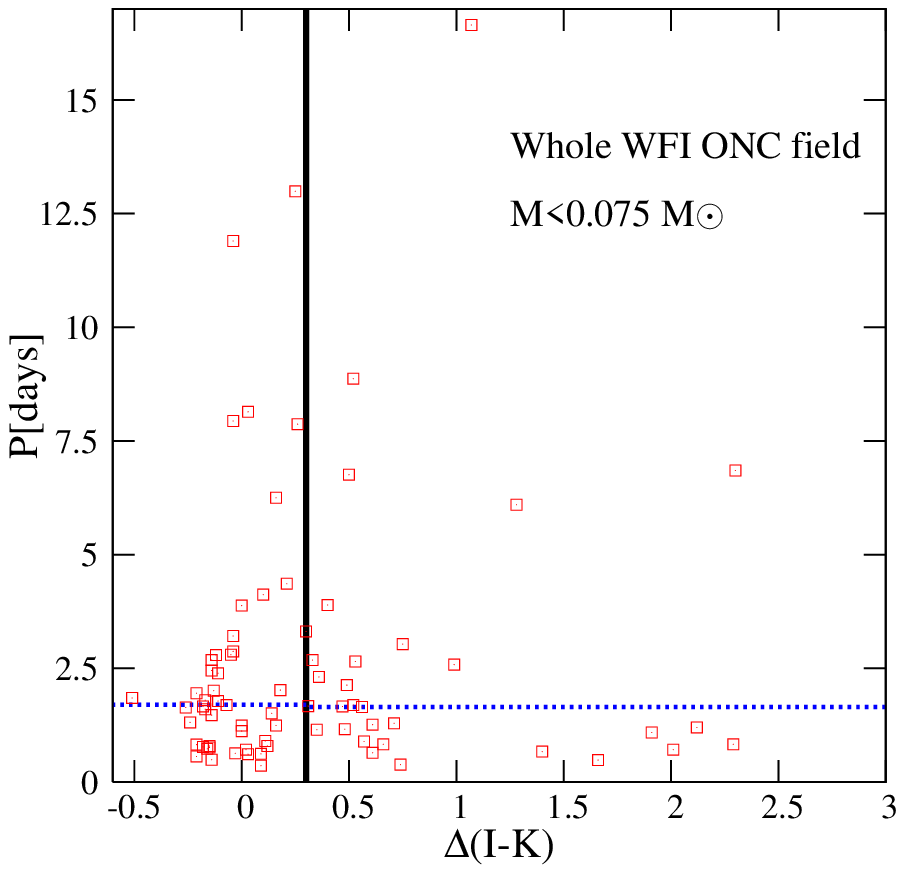}\\
\caption{Rotational period as a function of $\Delta$(I-K) for periodic variables with masses $>$0.4\,M$_{\odot}$, 0.075-0.4\,M$_{\odot}$, and $<$0.075\,M$_{\odot}$ for the whole WFI field. Vertical solid lines indicate $\Delta$(I-K)=\,0.3, with objects located right to this line showing an infrared excess. The median rotational periods for objects with and without an infrared excess are shown as horizontal dotted lines in all three mass regimes.}
\label{PvsQ}
\end{figure}
\begin{figure}
\centering
\includegraphics[width=8.5cm, height=16cm]{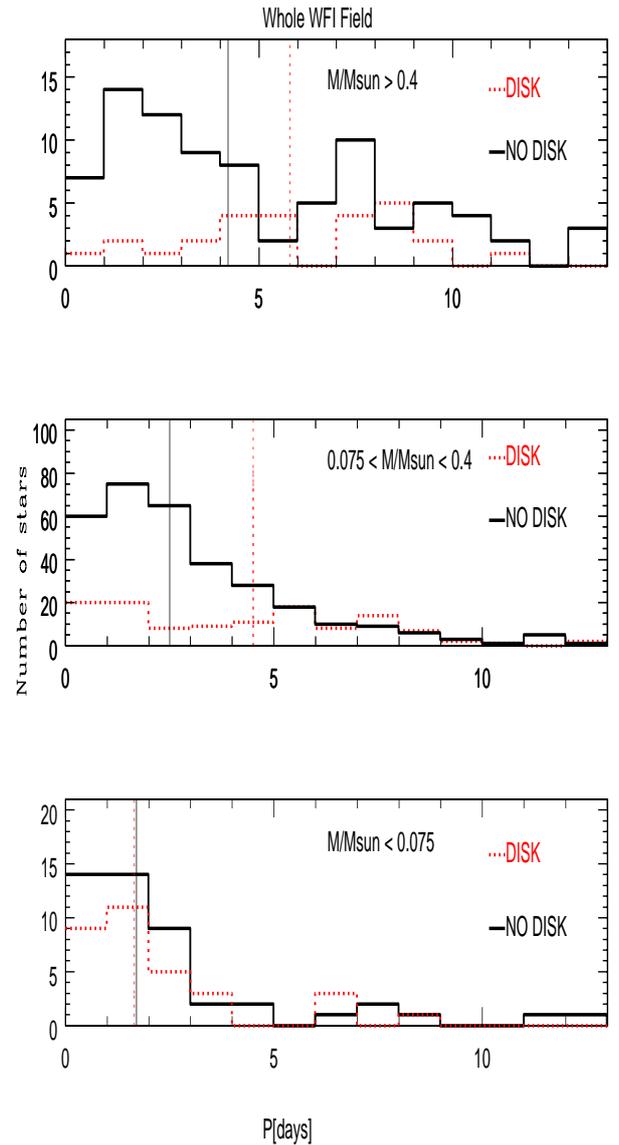}
\caption{Period distribution for objects with and without disks in the three mass bins investigated here (see text for details). Solid lines show the period distribution of objects without disks, while dotted lines represent the distribution for disk holders. Median rotational periods are shown as vertical lines.}
\label{Pmasshisto}
\end{figure}
\section{Correlation between infrared excess and rotational periods}\label{sec:PvsQ}
Below we investigate whether there is a correlation between the period distribution of the periodic variables measured in Paper\,1 and the infrared excess. As described in the introduction such a correlation was found in some studies for the higher mass periodic variables (e.g. H2002, Rebull et al. \cite{Rebull2006}, Cieza \& Baliber, \cite{Cieza2007}). Here we mainly want to find out whether there is a rotation-disk connection in the very low and substellar mass regime, and compare these results with higher mass objects.\\
Figure\,\ref{PvsQ} shows the rotational period as a function of the $\Delta$(I-K) index for objects in the whole WFI field. Since mass effects may bias any existing correlation between rotation and disks we perform all following analysis for the periodic variables in the already defined three mass bins. In the highest and intermediate mass regime, Fig.\,\ref{PvsQ} shows that fast rotators (i.e. $P\,\lesssim\,2\,days$) are mostly objects without NIR excess, while there is a significant fraction of slow rotators without indications of a disk, in addition to the slow rotators with NIR excesses detected. In the substellar regime no trend is clear from Fig.\,\ref{PvsQ}. The horizontal lines in each diagram indicate the median rotational periods for objects with and without IR excess (see Table\,3). In order to give statistical significance to the found trends we performed a two-sample K-S test, which shows differences in the period distribution of objects with and without disks in the highest mass bin at $>$\,2\,$\sigma$ level ($P$\,=\,0.02, $n_1$\,=\,27, $n_2$\,=\,85), in the intermediate mass bin at $>>$\,3\,$\sigma$ level ($P$\,=\,$10^{-6}$, $n_1$\,=\,121, $n_2$\,=\,324), while the period distribution of disks and non-disks holders in the substellar mass bin are statistically identical ($P$\,=\,0.95, $n_1$\,=\,33, $n_2$\,=\,48).\\
The period distribution of the periodic variables with and without disks in the three mass bins are easier to visualise in Fig. \,\ref{Pmasshisto}, where the red dotted histograms represent the period distributions of periodic variables with disks and the solid histograms are the corresponding period distributions for objects without disks. Figure\,\ref{Pmasshisto} shows that the distribution of objects with and without disks in the lowest mass bins look very similar. In the high and intermediate mass bins objects with disks rotate on average $\approx$\,1.4-1.8 times slower than objects without disks. In the substellar regime no such difference in the median rotational periods of objects with and without disks was found.\\
The same trend is found when the Q$_{JHHK}$ index is used as diagnosis of a circumstellar disk.\\
In addition to mass effects, age effects may bias the resulting correlations, since older objects tend to rotate faster and the disk-fraction among an older population should also be smaller. We suggested a possible age spread in the ONC in Paper\,1, with younger objects being located inside R$_\mathrm{cluster}$. Therefore it would in principle be interesting to do the analysis shown in Fig.\,\ref{Pmasshisto} for inside and outside R$_\mathrm{cluster}$. However, due to the reduced number of objects in the two spatial regimes and, particularly in the substellar mass regime, no statistically meaningful analysis could be obtained. We only show the median values of the rotational period distributions in Table\,3, which may suggest differences between the two spatial regions.\\
\begin{figure}
\centering
\includegraphics[width=8.1cm, height=6.5cm]{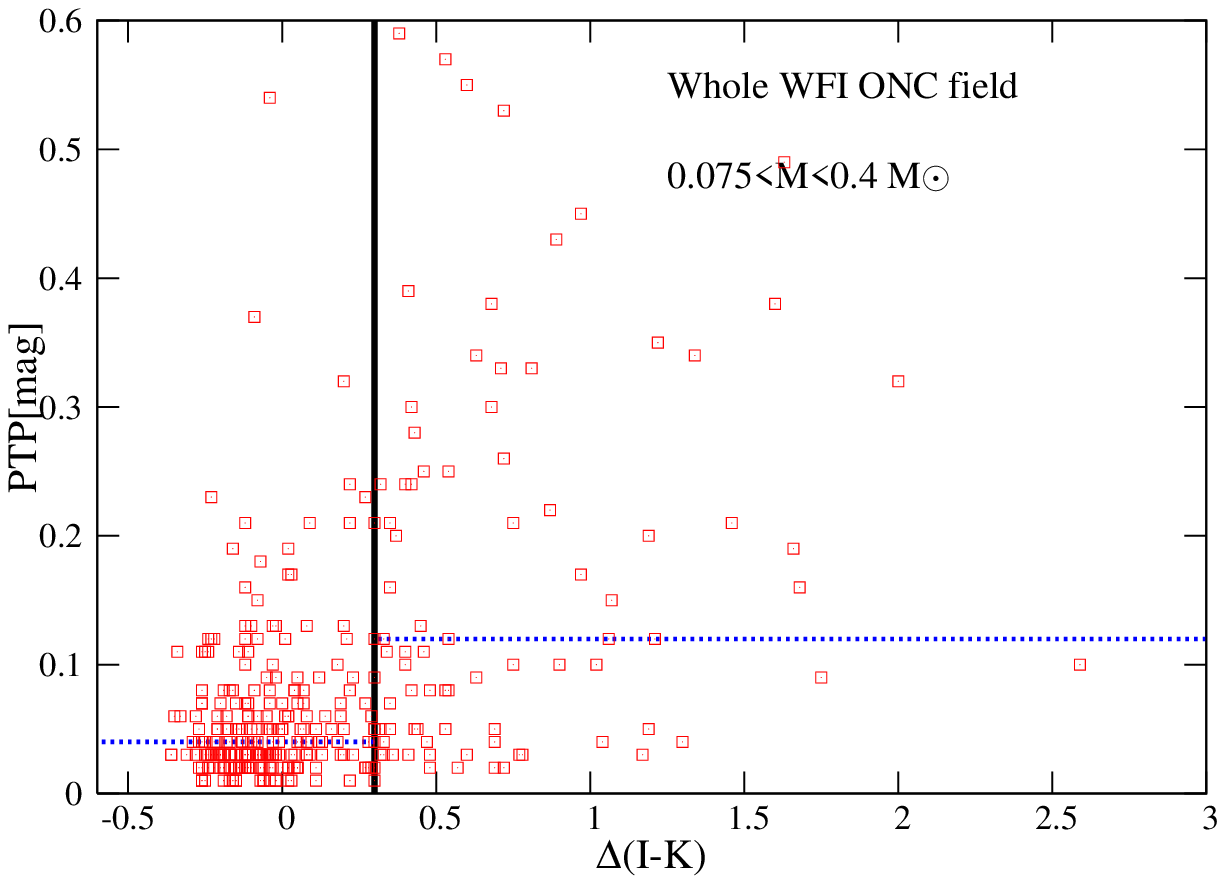}\\
\includegraphics[width=8.1cm, height=6.5cm]{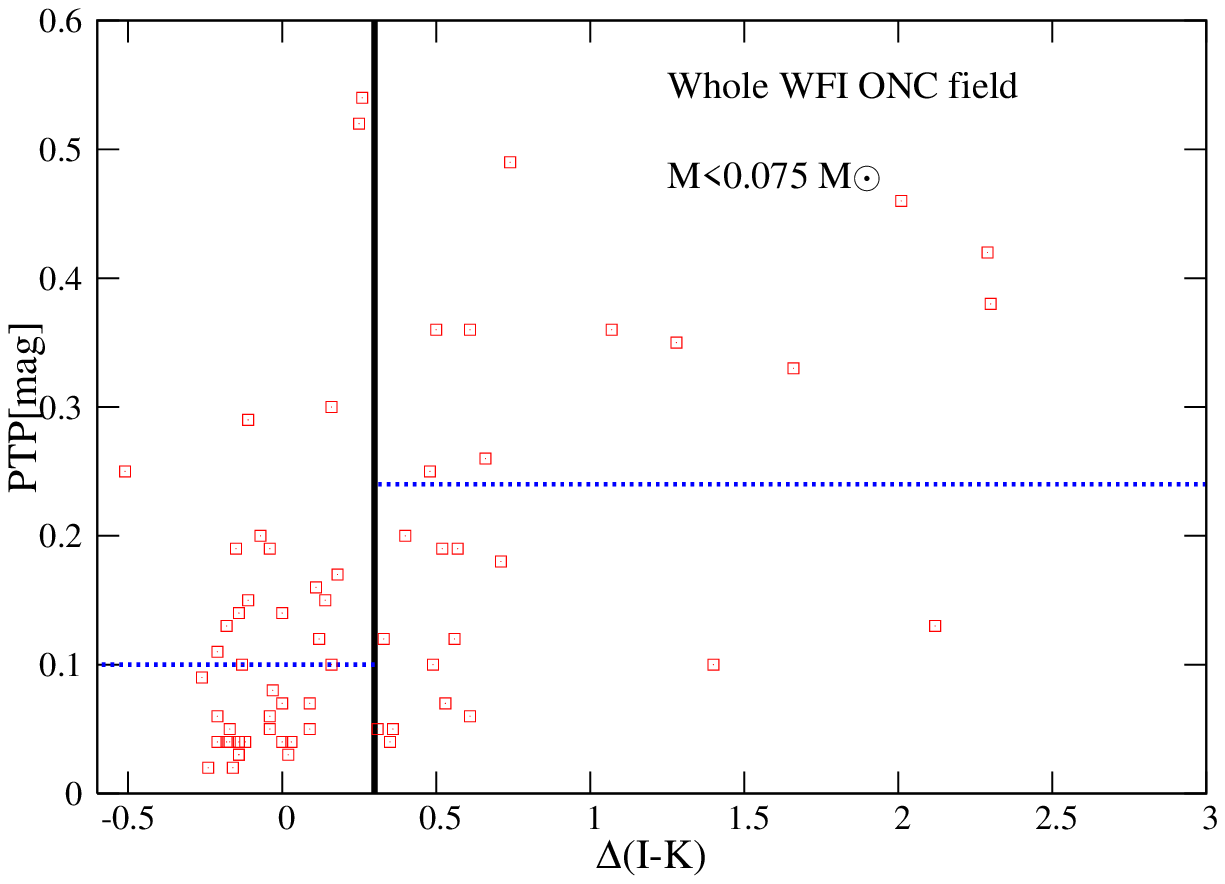}
\caption{The same as Fig.\,\ref{PvsQ} but for the peak-to-peak (ptp) amplitude.}
\label{PTPvsQ}
\end{figure}
\begin{figure}
\centering
\includegraphics[trim = 0mm 0mm 0mm 60mm,clip,width=9cm, height=11.5cm]{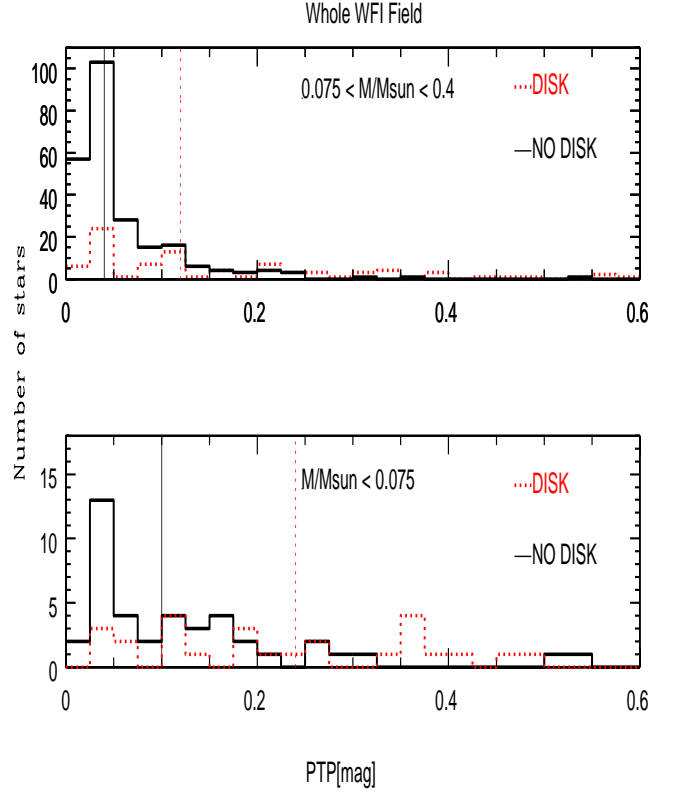}
\caption{Same as Fig.\,\ref{Pmasshisto} but for the peak-to-peak (ptp) amplitude distribution. Only the intermediate and substellar mass bin is shown since we have no information on the ptp amplitudes for most of the highest mass periodic variables from H2002.}
\label{PTPmasshisto}
\end{figure}

\section{Correlation between infrared excess and peak-to-peak amplitudes}

In the following we analyze whether there is a correlation between the peak-to-peak (ptp) amplitude of the periodic variables and NIR excess, i.e. the presence of a circumstellar disk. Since H2002 did not provide ptp amplitudes, we use here only the periodic variables measured in Paper\,1 and not the combined sample as in previous sections (for details in how the ptp amplitudes were determined see Paper\,1). Due to the small number of objects with M\,$>$\,0.4\,M$\mathrm{_{\odot}}$ we only investigate the ptp-disk correlation in the 0.4-0.075\,M$\mathrm{_{\odot}}$ and $<$\,0.075\,M$\mathrm{_{\odot}}$ mass bins.\\
Figure\,\ref{PTPvsQ} shows the ptp amplitudes as a function of the $\Delta$(I-K) index for these two mass intervals for the whole WFI field. The horizontal lines in each diagram indicate the median ptp amplitudes for objects with and without IR excess. From Fig.\,\ref{PTPvsQ} it is evident that objects with a NIR excess have on average about a factor 2-3 larger ptp amplitudes than those without a NIR excess. For the intermediate mass bin the two-sided K-S test gives a 10$^{-7}$ chance that the distributions of ptp amplitudes of objects showing infrared excess and the corresponding distributions of non-excess objects belong to the same population ($n_1$\,=\,93, $n_2$\,=\,243). In the substellar mass regime the K-S test implies a difference of the two populations at 3\,$\sigma$\, level ($P$\,=\,0.001, $n_1$\,=\,30, $n_2$\,=\,43). Among the objects with NIR excess there seems to be a trend of increasing ptp amplitude with increasing amount of NIR excess. Figure\,\ref{PTPmasshisto} helps to visualise the ptp distributions of objects with (dotted histograms) and without (solid histograms) disks. Median ptp values are shown in Table\,4. The lack of objects with ptp amplitudes below 0.025\,mag among the disk-holders is evident from both Figs.\,\ref{PTPvsQ} and \,\ref{PTPmasshisto}. We found that $\approx$\,79\,$\pm$\,2\,\% of all periodic variables without disks have ptp amplitudes below 0.1\,mag, while the corresponding fraction among the objects with disks is about 35\,$\pm$\,4\,\%. These results reveal a very robust indication of differences in the variability level of objects with and without circumstellar disks, in which objects with disks show on average much larger amplitude variations than objects without disks. \\
The same trend, median ptp values, and level of significance were found when the Q$_{JHHK}$ index is used as diagnosis of a circumstellar disk, providing additional support for the found correlation.\\
In Paper\,1 we found a correlation between the rotational period and the level of variability for periodic variables in the ONC and NGC2264. As stated in Paper\,1, this finding can probably be explained by differences in the spot coverage and/or differences in the magnetic field topologies. The strong rotation-disk and ptp-disk correlations found in the intermediate mass bin studied here may support this scenario. We found it rather interesting that the correlation between ptp amplitudes and NIR excess is strong in the substellar mass bin, while no rotation-disk correlation was found. This finding may indicate that even dipolar-like magnetic field topologies, which more likely produce large spot groups near the magnetic poles resulting in a high level periodic variability, may not be strong enough to allow effective disk coupling and braking in the substellar mass regime.
\begin{figure*}[t!]
\centering
\includegraphics[trim = 0mm 0mm 30mm 0mm,clip,width=6cm, height=5.5cm]{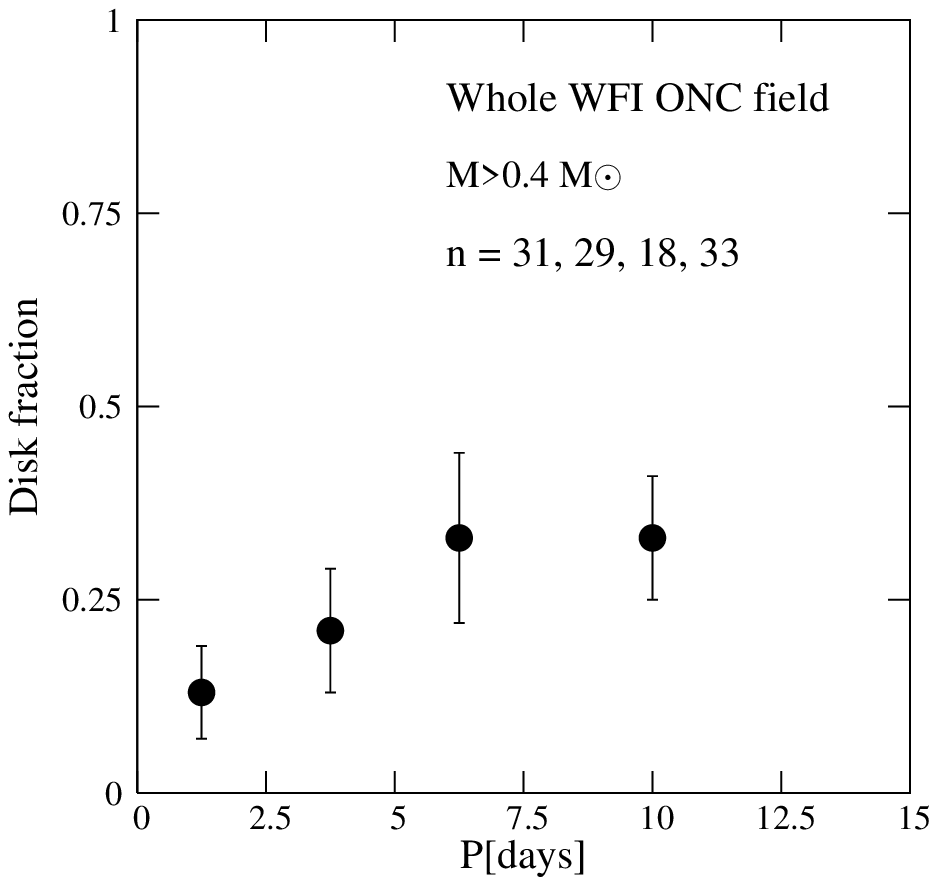}
\includegraphics[trim = 0mm 0mm 30mm 0mm,clip,width=6cm, height=5.5cm]{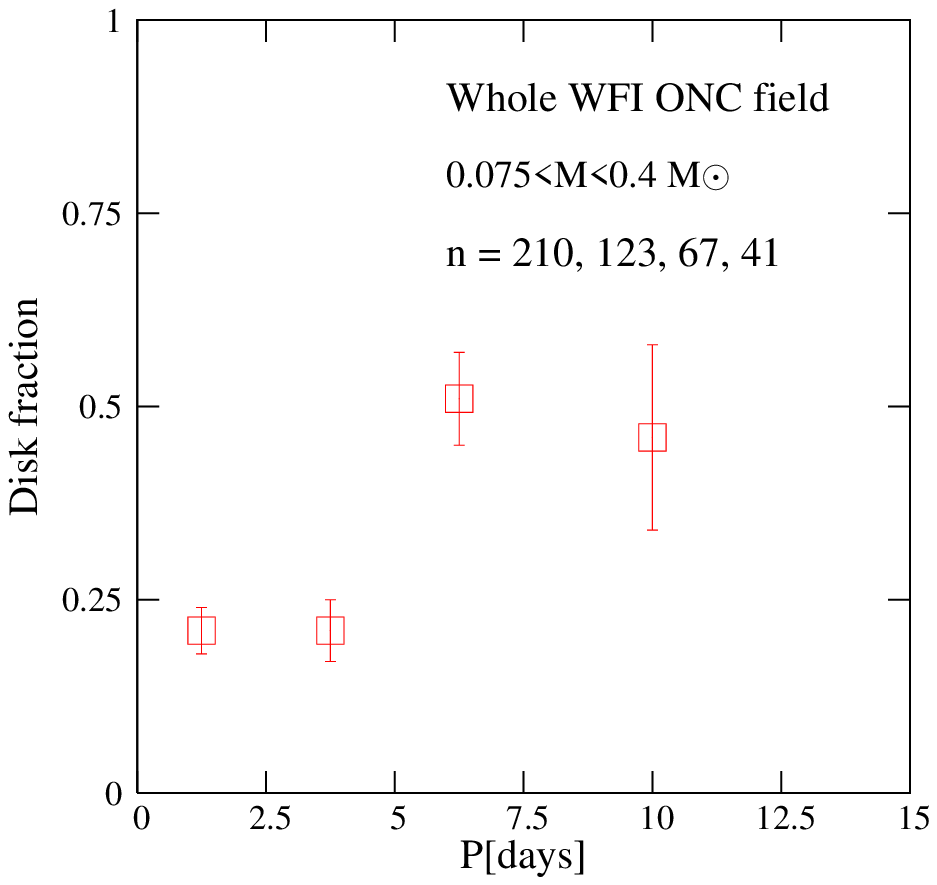}
\includegraphics[trim = 0mm 0mm 30mm 0mm,clip,width=6cm, height=5.5cm]{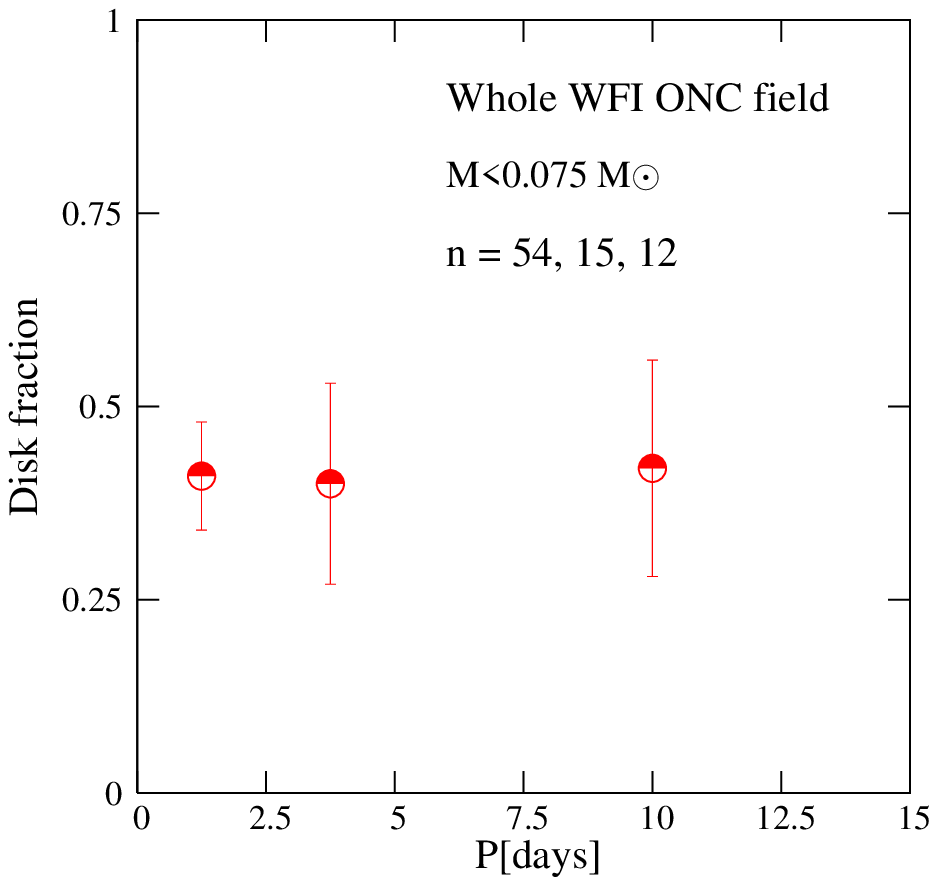}
\caption{Disk fraction as a function of rotational period for the three mass regimes studied here. The disk fractions are calculated for rotational period bins of 2.5 days with error bars representing the statistical standard error.}
\label{DFraction}
\end{figure*}
\begin{table*}
\centering
\begin{minipage}[c]{2\columnwidth}
\renewcommand{\arraystretch}{1.2}  
\caption{Median values of rotational period for objects with ($\Delta$(I-K)\,+) and without ($\Delta$(I-K)\,-) IR excess. }
\label{TabPTPvsQ}
\addtolength{\tabcolsep}{0.0pt}
\begin{tabular}{c|ccc|ccc|ccc|ccc|ccc|ccc} 
\hline 
\hline
& \multicolumn{6}{c|}{R$<$R$_\mathrm{cluster}$} & \multicolumn{6}{c}{R$>$R$_\mathrm{cluster}$}& \multicolumn{6}{|c}{Whole WFI field}\\
\hline
& \multicolumn{3}{c|}{$\Delta$(I-K)+} & \multicolumn{3}{c|}{$\Delta$(I-K)-}& \multicolumn{3}{c|}{$\Delta$(I-K)+} & \multicolumn{3}{c|}{$\Delta$(I-K)-}& \multicolumn{3}{c|}{$\Delta$(I-K)+} & \multicolumn{3}{|c}{$\Delta$(I-K)-}\\
\hline
Mass bins \footnote{Mass bins as defined in Sect.3: A corresponds to the bright bin containing stars with \textit{M} $\geq$ 0.4 \textit{M$\mathrm{_{\odot}}$}, B to the intermediate bin with 0.4$<$ M $<$ 0.075 \textit{M$\mathrm{_{\odot}}$}, and C to the faint bin in which all objects have \textit{M} $<$ 0.075 \textit{M$\mathrm{_{\odot}}$}.} & A & B& C & A & B& C & A & B& C & A & B& C& A & B& C & A & B& C \\  
\hline
N &13 & 56 &8 & 49 & 120 & 8 & 14& 65 & 25 & 36 & 204 & 40 & 27 & 121 & 33 & 85 & 324 & 48 \\
\hline
$<P>$[days] &6.3 & 4.7& 1.5 & 4.4 & 2.9 & 1.7 & 6.4 &3.8 &1.7& 3.8 & 2.0&1.6 &5.8 &4.5 &1.7&4.2&2.5&1.7\\
\hline
\end{tabular}
\renewcommand{\footnoterule}{} 
\end{minipage}
\end{table*}
\begin{table*}[t!]
\centering
\begin{minipage}[c]{2\columnwidth}
\renewcommand{\arraystretch}{1.2}  
\caption{Median values of ptp amplitudes for objects with ($\Delta$(I-K)\,+) and without ($\Delta$(I-K)\,-) IR excess.}
\addtolength{\tabcolsep}{3.7pt}
\begin{tabular}{c|cc|cc|cc|cc|cc|cc} 
\hline 
\hline
& \multicolumn{4}{c|}{R$<$R$_\mathrm{cluster}$} & \multicolumn{4}{c}{R$>$R$_\mathrm{cluster}$}& \multicolumn{4}{|c}{Whole WFI field}\\
\hline
& \multicolumn{2}{c|}{$\Delta$(I-K)+} & \multicolumn{2}{c|}{$\Delta$(I-K)-}& \multicolumn{2}{c|}{$\Delta$(I-K)+} & \multicolumn{2}{c|}{$\Delta$(I-K)-}& \multicolumn{2}{c|}{$\Delta$(I-K)+} & \multicolumn{2}{|c}{$\Delta$(I-K)-}\\
\hline
Mass bins \footnote{Mass bins as defined in Sect.3: A is not shown due to the small sample number, B corresponds to the intermediate bin with 0.4$<$ M $<$ 0.075 \textit{M$\mathrm{_{\odot}}$} and C to the faint bin in which all objects have \textit{M} $<$ 0.075 \textit{M$\mathrm{_{\odot}}$}.} & B& C & B& C & B& C & B& C & B& C & B& C \\  
\hline
N & 44 &8 & 97 & 8 & 48 & 22 & 150 & 35 & 93 & 30 & 243 & 43 \\
\hline
$<PTP>$[mag] &0.12 & 0.24& 0.05 & 0.16 & 0.11 &0.23 &0.03& 0.08 &0.12 &0.24&0.04&0.1\\
\hline
\end{tabular}
\renewcommand{\footnoterule}{} 
\end{minipage}
\end{table*}

\section{Disk frequency}
The disk frequency among our sample of 638 periodic variables can be estimated from the number of objects with a NIR excess. According to the $\Delta$(I-K) index 28\,$\pm$\,2\,\% of them have circumstellar disks. From the more conservative Q$_{JHHK}$ index, a lower disk frequency of about 20\% was derived. It has to be pointed out that both disk indicators are biased towards a small disk fraction since many periodic variables are WTTSs. If all stars in the ONC are considered the disk-frequency is substantially higher, namely between 50\% and 90\% (Hillenbrand et al. \cite{Hill98}). These numbers are consistent with the considerably larger fraction of irregular variables (CTTSs) relative to periodic variables we found in Paper\,1 for the ONC (ratio of irregular to periodic variables $\approx$\,1.7). In addition, as will be discussed in Sect.\,\ref{Discussion}, the derived disk fraction may also be smaller than expected since NIR excess is not an optimum disk indicator.\\
In order to investigate a possible dependence of the disk fraction on the mass of the periodic variable, we derived the fraction of objects with and without disks in the three mass bins investigated here. We found that the disk frequency slightly increases from 24\%\,$\pm$\,4\% to 27\%\,$\pm$\,2\% for the highest to the intermediate mass bins, while the increase in the disk fraction among the substellar objects is stronger (41\%\,$\pm$\,5\%)\footnote{Disk fraction uncertainties correspond to the statistical standard error $[(diskfraction*(1\,-\,diskfraction)/\,N)]^{1/2}$.}. Figure\,\ref{DFraction} shows the disk fraction as a function of rotational period, in the three mass regimes studied. In the high and intermediate mass bin, the disk fractions are computed in bins of 2.5\,days up to 7.5 days, while due to the small number of slow rotators, the last bin includes all objects with a rotational period larger than 7.5\,days. In the substellar regime, due to the much smaller sample numbers, only three bins are used, i.e. P\,$<$\,2.5\,days, 2.5\,$\leq$\,P\,$<$\,5\,days, and P\,$\geq$\,5\,days. From Fig.\,\ref{DFraction} is evident that there is a trend towards larger disk fractions for the slower rotating stellar objects, while in the substellar mass regime no such trend is indicated at all, despite the fact that the disk fraction is rather high for all periods. Figure\,\ref{DFraction} fully supports the finding of a correlation between period and NIR excess for objects with masses $>$\,0.075\,M$\mathrm{_{\odot}}$.\\
In each mass regime we found that the disk frequency is always higher inside R$_\mathrm{cluster}$ than in the presumably older region outside R$_\mathrm{cluster}$.
\section{Discussion \label{Discussion}}
\subsection{Higher mass objects}

According to Sect.\,\ref{PvsQ}, a correlation between rotational periods and the presence of circumstellar disks was found in the higher mass regime at $>$\,2\,$\sigma$ level. As outlined in the Introduction, in the past decade several authors have investigated correlations between rotation rates and the presence of circumstellar disks in the solar and subsolar mass regime (i.e. M\,$\gtrsim$\,0.2\,M$_{\odot}$; e.g. H2002; Makidon, \cite{Makidon2004}; Lamm et al. \cite{Lamm05}; Rebull et al. \cite{Rebull2006}; Cieza and Baliber, \cite{Cieza2007}; Nguyen et al. \cite{Nguyen2009}). Some studies found evidence for such a connection and others did not. Most of the discrepant results are probably related to the inhomogeneous samples that have been used for comparison. Due to the strong dependence of the rotational period with mass and age, proper comparisons can only be done if the samples contain periodic variables in the same mass regime and of similar ages. In the ONC, the mass effect was taken into account in the work of Cieza \& Baliber (\cite{Cieza2007}). They use IRAC excesses as disk indicators together with rotational periods from the literature, and find for objects with masses of $\gtrapprox$\,0.4\,M$_{\odot}$ clear evidence of angular momentum regulation by circumstellar disks, confirming the correlation already found by H2002.\\
H2002 argued for a significant correlation between $\Delta$(I-K) and rotational periods for objects with masses of $\gtrsim$\,0.2\,M$_{\odot}$ (with very few objects with M\,$\lesssim$\,0.2\,M$_{\odot}$) in the ONC. They tested the significance of the mentioned correlation by a non-parametric Spearman test, which resulted in less than a 10$^{-10}$ ($n$\,=\,325) chance that the quantities are not correlated. Since they did not provide the significance of the mentioned correlation for different mass regimes, we reanalyzed their data set and found that when dividing in a higher and a lower mass bin (i.e. M\,$\lesseqgtr$\,0.4\,M$_{\odot}$ or M\,$\lesseqgtr$\,0.25\,M$_{\odot}$ with the D'Antona and Mazzitelli (\cite{Dantona97}) models used in H2002), the level of significance of the NIR-rotation correlation decreases substantially to 2x10$^{-3}$ ($n$\,=\,157) for objects with M\,$>$\,0.4\,M$_{\odot}$ and 1x10$^{-6}$ ($n$\,=\,168) for the lower mass objects in their sample. The correlation is much stronger for the less massive objects in their sample, which agrees with our results.\\
The same test performed for our estimated $\Delta$(I-K) index resulted in a 10$^{-9}$ ($n$\,=\,638), 10$^{-2}$ ($n$\,=\,112), 10$^{-8}$ ($n$\,=\,445), and 6x10$^{-1}$ ($n$\,=\,81) chance that NIR-excess from the $\Delta$(I-K) index and rotational periods are not correlated for the whole sample, the higher, intermediate, and lower mass objects, respectively. These results show a trend which is consistent with H2002, in the sense that both higher mass objects and substellar objects tend to have a weaker correlation than the intermediate mass objects in our sample.\\ 
We emphasise that it is not at all the purpose of this paper to investigate the degree of significance of a NIR excess-rotation correlation for higher mass stars ($>$ 0.4\,M$_{\odot}$), but mostly to investigate the relative change of such a correlation with mass. This relative change is less affected by the use of a non-ideal disk indicator, because the same disk indicator is used in all investigated mass regimes.

\subsection{Substellar mass objects \label{Discussion1}}

According to Figs.\,\ref{PvsQ} and \,\ref{Pmasshisto} there is no correlation between the NIR excess (i.e. the presence of a circumstellar disk) and the rotation rate in the substellar mass regime. The lack of this correlation can in principle be caused by the combination of the following effects: 
\begin{enumerate}
\item NIR excesses are not sufficiently good accretion disk indicators.
\item There is a certain angular momentum loss in the BD mass regime at the age of $\approx$\,1\,Myr, but the following effects strongly weaken any correlation:
\begin{description}
 \item - vast spread in rotational periods.
 \item - sample has a too large spread in mass and age.
\end{description}
\item Angular momentum loss rates are indeed significantly lower in BDs than in higher mass objects.
\end{enumerate}
Below we will discuss the possible relevance of these effects in more detail .

\subsubsection{Are NIR excesses good accretion disks indicators?}
As outlined in the Introduction, theoretical models require ongoing mass accretion from a circumstellar disk onto the star to account for angular momentum losses. To test these theoretical predictions observationally it is necessary to measure mass accretion rates. Some of the more commonly used mass accretion indicators are the equivalent width (EW) of H$\alpha$ or other suitable emission lines (e.g. Ca II triplet), continuum veiling, and UV excess. Near-infrared excesses can only account for the presence of dust in a circumstellar disk which does not imply that gas is being accreted from this disk onto the star. This means it is possible to detect a NIR excess from a purely irradiated circumstellar disk even if the mass accretion rate is zero (see for example Fig.\,6 in Hillenbrand et al. \cite{Hill98}). Nevertheless, an impressive correlation between NIR excess and mass accretion indicators such as the H$\alpha$ emission line width has been found for many YSOs over a broad mass range (e.g. Sicilia-Aguilar et al. \cite{Sicilia2005}, \cite{Sicilia2006}). This strong correlation allows us to use NIR excess as a good accretion disk indicator in this study, since a more direct accretion indicator (e.g. H$\alpha$ emission line flux) is not available for our sample.\\
Near-infrared excesses have been widely used as diagnostic of a circumstellar disk and have succeeded in detecting large fraction of very low mass stars and brown dwarfs with circumstellar disks (e.g. Lada et al. \cite{Lada2000}; Muench et al. \cite{Muench2001}). Muench et al. (\cite{Muench2001}) found that about 50\% of their brown dwarf candidates in the Trapezium region show NIR excesses suggestive of a circumstellar disk. This fraction is similar to the one found by Lada et al. (\cite{Lada2000}) for very low mass objects, which was increased to about 85\% when L-band (3.8 $\mu$m) observations were used. A similar increase in the disk fraction among the brown dwarf population studied here would therefore be expected from longer wavelength observations. In the last few years, IRAC excesses were used in young clusters to unambiguously determine the presence of a circumstellar disk (e.g. Megeath et al. \cite{Megeath2005}). The problem with IRAC data arises in the inner region of the ONC, where the nebular background and the stellar density is very high, and therefore a considerable number of even bright objects are missing IRAC detections (Megeath et al. \cite{Megeath2005}, Rebull et al. \cite{Rebull2006}). Up to now, only objects in the ONC with spectral types earlier than M2 can be precisely analyzed and searched for disk presence and a rotational-disk connection by means of IRAC data (Rebull et al. \cite{Rebull2006} and Cieza \& Baliber \cite{Cieza2007}). We then have to rely on the less favourable NIR excesses as indicators of disks around very low mass and substellar objects.\\
We like to remind the reader that the photometric method used for measuring periodic brightness modulations is more efficient among the WTTSs than for CTTSs, since irregular variations can add a strong noise to the periodic modulation, preventing the detection of the periodic signal. Due to this noise we are definitely missing periodic variables among the young highly active CTTSs, which are expected to have larger amounts of NIR excess (i.e. larger mass accretion rates) than the more common WTTSs in our sample of periodic variables. This bias is an additional factor which prevents us from finding a stronger correlation between rotational periods and the presence of a circumstellar disk.\\
In summary, we believe that the used NIR disk indicator is not a major cause for the lack of a disk-rotation correlation found in the substellar regime. 

\subsubsection{Correlation weakening/destroying effects}

Below we discuss the effects that can weaken or even destroy a correlation between rotation and mass accretion, even in the case of an ``ideal'' mass accretion indicator.
\begin{itemize}
\item \textit{Large spread of rotational periods}:\\ From Fig.\,\ref{PvsQ} it is clear that the period distribution of the periodic variables in all three mass bins show a vast spread. If this spread can weaken or even destroy a correlation between rotation and NIR excess, it should in principle affect all three mass bins. To test whether there is a larger spread in rotational periods in the BD regime than in the intermediate mass regime, in which a highly significant correlation was found, we computed $P/P_{median}$ values and analyzed the median and standard deviation ($\sigma$) of these $P/P_{median}$ distributions. In addition, a \textit{log-normal} distribution was also fitted and both the geometric medians and the multiplicative $\sigma$ were compared. The $\sigma$ values provide the width of the distribution and therefore give information on the spread of the values. We found that the distribution of $P/P_{median}$ in the substellar mass regime is broader than the corresponding distribution of the intermediate mass objects (i.e. the $\sigma$ values are a factor $\approx$\,1.6 larger). In addition, even when the distribution is broader, the number of outliers\footnote{Number of objects with values outside the following range: Q$_1$-1.5x(Q$_3$-Q$_1$) and Q$_3$+1.5x(Q$_3$-Q$_1$), where Q$_1$ and Q$_3$ are the first and third quartiles. For our sample, all outliers have values higher than the upper limit, i.e. longer periods.} is higher in the BD mass regime than in the intermediate mass regime, with 14\% and 3\% of outliers in each region, respectively. \\
The larger spread in rotational periods among the BDs in our sample seems to be an important correlation decreasing factor in the BD regime.\\
\item \textit{Mass and age effects}:\\ As stated in Sect.8.1, the correlation between rotational period and NIR excess can be biased by the strong dependence of rotation with mass and age.\\
If a sample of periodic variables with a broad mass range is used in order to find a correlation with NIR excess the following effect may mimic a rotation-disk connection: higher mass objects rotate on average slower and show on average larger NIR excesses than lower mass objects, the latter due to the higher contrast between the photosphere and the inner disk emission (e.g. Littlefair et al. \cite{Littlefair2005}). \\
An age spread may also mimic an intrinsic correlation between rotational periods and disks. \\ 
The intermediate mass regime has a mass range of about 0.4/0.075\,$\approx$\,5. However, the mass range in the BD regime is most likely much smaller since we have only very few objects below 0.025\,M$_{\odot}$, i.e. the mass range is there about 3 (0.075/0.025). Due to the smaller mass range in the BD regime it cannot be excluded that this effect contributes to the lack of a correlation between disks and rotation rates. To test this hypothesis, we divided the intermediate mass regime in two mass bins, namely objects with 0.4$<$\,M\,$<$ 0.17\,M$_{\odot}$ (intermediate-high) and 0.17$<$\,M\,$<$ 0.075\,M$_{\odot}$ (intermediate-low), and analyzed the new period distributions of objects with and without NIR excess. The KS test resulted in a 99\% and 92\% probability that the period distributions of objects with and without NIR excess belong to different populations in the intermediate-high and intermediate-low mass bins, respectively. The mass ranges in both these newly defined mass bins are about 2.3 and therefore smaller than the mass range in the substellar regime; nevertheless, rotation-disk correlations are found to be more significant than in the substellar mass regime. The statistical significance of the mentioned correlation decreases toward lower masses, which most probably indicates that mass effects are not relevant within the mass intervals studied here.\\
\end{itemize}

\subsubsection{Are angular momentum loss rates smaller in BDs than in higher mass objects?}

We discuss here possible physical processes which can account for smaller angular momentum loss rates in the substellar mass regime compared to higher mass objects.

\begin{itemize}
\item \textit{Low ionization}:\\All hydro-magnetically driven angular momentum loss mechanisms require a sufficiently high ionization of the disk and/or wind involved in the angular momentum transport from the star and/or disk to this circumstellar matter. Without a sufficient ionization the magnetic field would only poorly couple to the circumstellar matter, and therefore mass accretion rates and angular momentum transport would be correspondingly smaller. Since the X ray and UV fluxes of BDs relative to their bolometric fluxes (i.e. $L_x/L_{Bol}$) are certainly much lower than for higher mass stars, it is quite conceivable that these effects reduce the angular momentum braking in the BD regime. This low ionization/magnetic coupling consideration can presumably explain the strong decrease of $\dot{M}$ with decreasing $M_{star}$ discussed in the next section.\\
\item \textit{$\dot{M}$ versus $M_{star}$ relation \label{MassAc vs Mass}}:\\Observations over a broad mass range from intermediate mass stars down to brown dwarfs reveal a very strong dependence of the mass accretion rate on stellar mass with the global trend $\dot{M}$\,$\propto$\,$M_{star}^2$ (Hartmann et al. \cite{Hartmann06}). An even stronger dependence, $\dot{M}$\,$\propto$\,$M_{star}^{2.8}$ has been reported in the subsolar mass regime down to the substellar limit by Fang et al. (2009). This dependence explains the very small mass accretion rates of $\lesssim$\,10$^{-10}$\,M$_{\odot}$ measured in very low mass and substellar objects (e.g. Muzerolle et al. 2005). Obviously this non linear relation between mass accretion and stellar mass indicates that the relative braking $\dot{J}/J$ is strongly mass dependent, with a substantially smaller braking in the lower mass objects.\\
In addition, Hartmann et al. (\cite{Hartmann06}) have shown that $d\dot{M}/dt$ is probably also highly mass dependent in the sense that it strongly increases with decreasing mass; i.e. $\dot{M}$ decreases much faster with evolutionary age in BDs than in solar mass stars.\\
\item \textit{Relevant time scales for disk-locking: $\tau_{lock}$ and $\tau_{D}$}:\\
If one assumes that the disk locking scenario explains the rotational data one can consider two extreme cases: case 1; all objects are locked to their disks at their birth, and the amount of angular momentum loss strongly depends on how long this locking is effective (i.e. how long $\tau_{lock}$) is, and case 2; they are not locked to their disks at early stages and it takes a certain time $\tau_{D}$ until they are locked. \\
In the first case, the above discussed $\dot{M}/dt$ versus mass consideration do indeed argue that $\tau_{lock}$ in BDs is much shorter than in solar-mass stars, since in the former case $\dot{M}$ will decrease much more rapidly. Therefore BDs are expected to be braked for a shorter time than higher mass objects, after which they spin up conserving angular momentum or losing angular momentum in a moderate way.\\
Concerning case 2, Hartmann (\cite{Hartmann}) argued that disk-locking is not instantaneous but a certain time $\tau_{D}$ is needed before disk-locking is achieved. Equation \ref{Tau} (which is Eq.\,8 in Hartmann \cite{Hartmann}) shows that this time is proportional to the mass, the angular velocity, and to the inverse of the accretion rate: 
\begin{equation}
\tau_D\,\gtrapprox\,4.5x10^6\, yr\, f\, \frac{M_{0.5}}{\dot{M}^{-8}},
\label{Tau}
\end{equation}
in which M$_{0.5}$ and $\dot{M}^{-8}$ are the mass and mass accretion rates in units of 0.5\,M$_{\odot}$ and 10$^{-8}$\,M$_{\odot}$yr$^{-1}$, respectively and \textit{f} is the angular velocity in units of the breakup velocity.\\
Typical values of these quantities for a 0.5\,M$_{\odot}$ star are $\dot{M}$=10$^{-8}$\,M$_{\odot}$yr$^{-1}$ and f\,=\,0.1 (see Herbst et al. \cite{H2001}), which result in a $\tau_{D}$\,$\approx$\,0.5\,Myr. Since $\dot{M}$ in BDs is highly time dependent $\tau_{D}$ will strongly depend on the evolutionary phase. If one assumes an optically observable $\approx$\,1\,Myr old BD, a typical $\dot{M}$\,=\,5x10$^{-10}$\,M$_{\odot}$yr$^{-1}$ is expected. In this case $\tau_{D}$ is about 10\,Myr (for f\,$\approx$\,0.7 which is larger in BDs because they rotate close to break up velocity). 
This all means that disk coupling in BDs can only be achieved within in a reasonable time scale in very early evolutionary phases, when $\dot{M}$ was presumably two to three orders of magnitude higher.\\

\end{itemize}

\subsubsection{Concluding remarks}
There is strong observational evidence that the mass accretion rate $\dot{M}$ for low mass stars and substellar objects is highly mass dependent ($\dot{M}$\,$\varpropto$\,$M^{2\,-\,2.8}_{star}$), and therefore one would expect that the ratio $\dot{J}/J$ depends on mass in a similar manner. With these dependencies it is not surprising that no correlation between the NIR excess and the rotation rate was found in the substellar regime.\\
Before reliable conclusions can be drawn, independent disk indicators (e.g. IR excess at longer wavelengths) should be correlated with rotation and, if possible, future substellar samples should include objects at slightly different ages.

\section{Summary}

We correlated rotational periods and peak-to-peak (ptp) amplitudes with NIR excess in a large sample of VLM stars and BDs in the ONC. Our results are summarized in the following:
\begin{enumerate}
\item We analyzed 638 periodic variables in the ONC from Paper\,1 and H2002 with IJHK measurements.
\item We estimated masses from the J\,versus\,I-J color-magnitude diagram and found 81 objects with masses in the substellar mass regime. 
\item We calculated the amount of NIR excess from a circumstellar disk by means of the $\Delta$(I-K) index and and, for comparison purposes, we also computed an extinction-free index Q$_{JHHK}$.  
\item We performed all our analysis in three mass bins, namely objects with masses $>$\,0.4\,$M_{\odot}$, 0.4-0.075\,$M_{\odot}$, and $<$\,0.075\,$M_{\odot}$. Our results show the existence of a highly significant rotation-disk correlation for the intermediate mass objects, in which objects with NIR excess tend to rotate slower than objects without NIR excess. In the highest mass bin there is a correlation at $>$\,2\,$\sigma$ level. Interestingly, and most important for this study, we found no correlation in the substellar regime.
\item We found a highly significant correlation between the ptp amplitudes and the NIR excess in the two mass bins (0.4-0.075\,$M_{\odot}$, and $<$\,0.075\,$M_{\odot}$) with available ptp values, in which objects with circumstellar disks show on average a factor two to three larger ptp amplitudes than those without indications of a disk. 
\item We estimated a disk frequency between 20\% and 28\% among the 638 periodic variables in the ONC. This fraction is smaller than the one estimated for all objects in the ONC by Hillenbrand et al. (\cite{Hill98}), which is expected because (\textit{a}) the photometric method used to derived rotational periods is more effective among the WTTSs than for the CTTSs, with the former group having a much smaller disk frequency, (\textit{b}) NIR excesses miss a certain fraction of objects with circumstellar disks.
\item The three mass regimes studied show different rotation-disk behaviors which may be attributed to the different physical properties of objects in each mass bin. \\
The first division line (0.4\,$M_{\odot}$) coincides with the onset of full convection. It may be possible that the different behavior between high and intermediate mass objects in our sample are related to the different ways magnetic fields are produced, resulting in different topologies, spots coverages and/or magnetic field strengths. \\
The second boundary is the substellar limit. The strong dependence of mass accretion rate on mass seems to be crucial to understand why the lower mass objects in our sample suffer smaller angular momentum losses than those with larger masses. Our observations suggest that at an age of about 1\,Myr disk-braking does not influence the rotational rate of most substellar objects.\\
In order to exclude possible observational biases by the method used here, future studies should consider IR excess at longer wavelengths or a direct mass accretion indicator.

\end{enumerate}
\begin{acknowledgements}
The authors thank Mark MacCaughrean for providing the VLT-ISAAC NIR data used for 113 objects. We especially thank the anonymous referee for her/his valuable comments on the original version of this manuscript. \\
M.V.R.L. acknowledges support of the International Max Planck Reseach School for Astronomy and Cosmic Physics of the University of Heidelberg.\\
This publication makes use of data products from the Two Micron All Sky Survey, which is a joint project of the University of Massachusetts and the Infrared Processing and Analysis Center/California Institute of Technology, funded by the National Aeronautics and Space Administration and the National Science Foundation. This work is based in part on data obtained as part of the UKIRT Infrared Deep Sky Survey.
\end{acknowledgements}

\end{document}